\documentclass[10pt]{article}
\usepackage{graphicx}

\def\Title#1{\begin{center} {\Large #1 } \end{center}}
\def\Author#1{\begin{center}{ \sc #1} \end{center}}
\def\Address#1{\begin{center}{ \it #1} \end{center}}

\newcommand\pubblock{\rightline{\begin{tabular}{l} Proceedings of the Second Annual LHCP\\ \pubnumber\\
         \pubdate  \end{tabular}}}

\newenvironment{Abstract}{\begin{quotation} \begin{center} 
             \large ABSTRACT \end{center}\bigskip 
      \begin{center}\begin{large}}{\end{large}\end{center} \end{quotation}}

\newenvironment{Presented}{\begin{quotation} \begin{center} 
             PRESENTED AT\end{center}\bigskip 
      \begin{center}\begin{large}}{\end{large}\end{center} \end{quotation}}

\def\beq{\begin{equation}}
\def\eeq#1{\label{#1}\end{equation}}
\def\eeqn{\end{equation}}

\def\beqa{\begin{eqnarray}}
\def\eeqa#1{\label{#1}\end{eqnarray}}
\def\eeqan{\end{eqnarray}}

\let\bar=\overbar

\def\Dslash{\not{\hbox{\kern-4pt $D$}}}
\def\dslash{\not{\hbox{\kern-2pt $\del$}}}

\def\msb{{\bar{\ssstyle M \kern -1pt S}}}

\usepackage{color}

\newcommand\pubnumber{ ATL-PHYS-PROC-2014-099 }

\newcommand\pubdate{\today}

\def\affiliation{
On behalf of the ATLAS Experiment, \\
Institute of Physics, University of Belgrade, Belgrade, Serbia\\
LAL, CNRS/IN2P3, Orsay Cedex, France}

\begin{document}
\large
\begin{titlepage}
\pubblock

\vfill
\Title{ Search for squarks and gluinos with the ATLAS detector in final states with jets and missing transverse momentum using 20.3 $fb^{-1}$ of $\sqrt{s}$ = 8 TeV proton-proton collision data  }
\vfill

\Author{ MARJANOVI{\'C} MARIJA }
\Address{\affiliation}
\vfill
\begin{Abstract}

Weak scale supersymmetry is one of the best motivated and studied Standard Model extensions.
It predicts the existence of new heavy coloured particles called squarks and gluinos which are the supersymmetric partners of the quarks and gluons, respectively.
The poster summarises results on inclusive searches for supersymmetric squarks and gluinos in events containing jets and missing transverse momentum without leptons.
The searches use the full data sample recorded in 2012 at $\sqrt{s}$=8~TeV centre-of-mass energy by the ATLAS experiment at the LHC.

\end{Abstract}
\vfill

\begin{Presented}
The Second Annual Conference\\
 on Large Hadron Collider Physics \\
Columbia University, New York, U.S.A \\ 
June 2-7, 2014
\end{Presented}
\vfill
\end{titlepage}
\def\thefootnote{\fnsymbol{footnote}}
\setcounter{footnote}{0}
%

\normalsize 


\section{Introduction}

Many SUSY models predict squarks $\tilde{q}$ and gluinos $\tilde{g}$ \cite{susy1}~\cite{susy2}(SUSY partners of quarks and gluons) that could be accessible at the LHC, thanks to large production cross-sections.
The partners of the neutral and charged Standard Model (SM) gauge and Higgs bosons are, respectively, the neutralinos ($\tilde{\chi}_i^0$) and charginos ($\tilde{\chi}_i^{\pm}$).
This analysis \cite{atlas}~\cite{0lep} is searching for supersymmetric particles assuming R-parity conservation (therefore $\tilde{q}$ and $\tilde{g}$ must be produced in pairs ($\tilde{g}\tilde{g}$, $\tilde{q}\tilde{q}$, $\tilde{q}\tilde{g}$)) and that the $\tilde{\chi}_1^0$ is the Lightest Supersymmetric Particle (LSP).
This search is aimed at final states containing at least 2 to 6 jets and large missing transverse energy ($E_T^{\mathrm{miss}}$).
In addition, a veto on leptons (electrons and muons) is applied (to be orthogonal to other ATLAS searches).
The main results of this analysis are relevant for constraining many models of new physics. 

\section{Signal region selection}

This analysis uses a combined jet + $E_T^{\mathrm{miss}}$ trigger and rejects any event failing to satisfy quality selection criteria designed to suppress detector noise and non-collision backgrounds.
To suppress W+jets and $t\bar{t}$ background a veto on isolated electrons and muons is applied.
As we are assuming the LSP is neutral and weakly interacting, only events with  $E_T^{\mathrm{miss}} > $ 160 GeV are kept.
Because of the high mass scale expected for the SUSY signal, the `effective mass', $m_{\mathrm{eff}} = \sum_{p_T > 40~GeV} p_T^{jets} + E_T^{\mathrm{miss}}$, is a powerful discriminant between the signal and most SM backgrounds.

\begin{figure}[htb]
\centering
\includegraphics[height=1.45in]{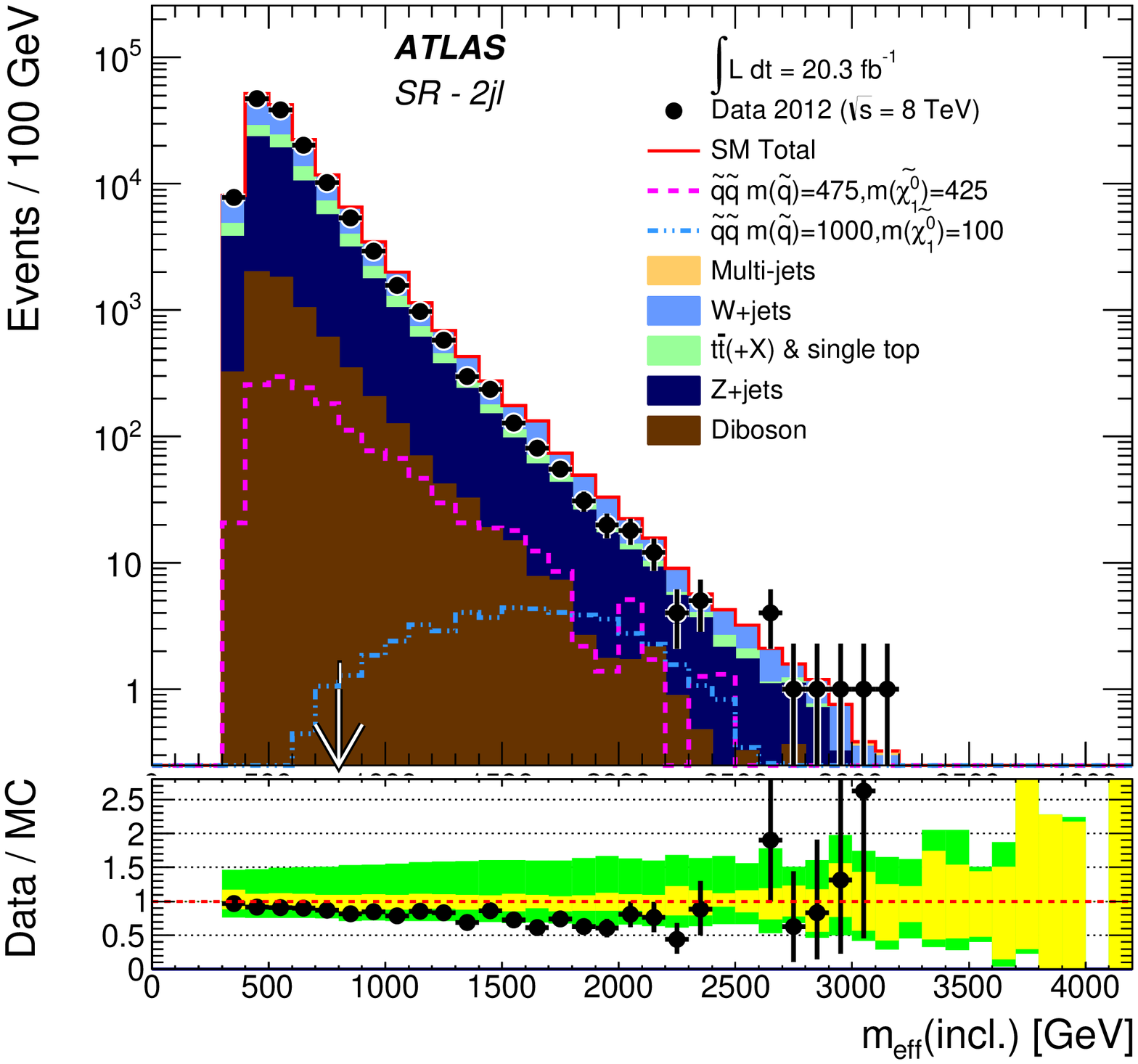}
\includegraphics[height=1.45in]{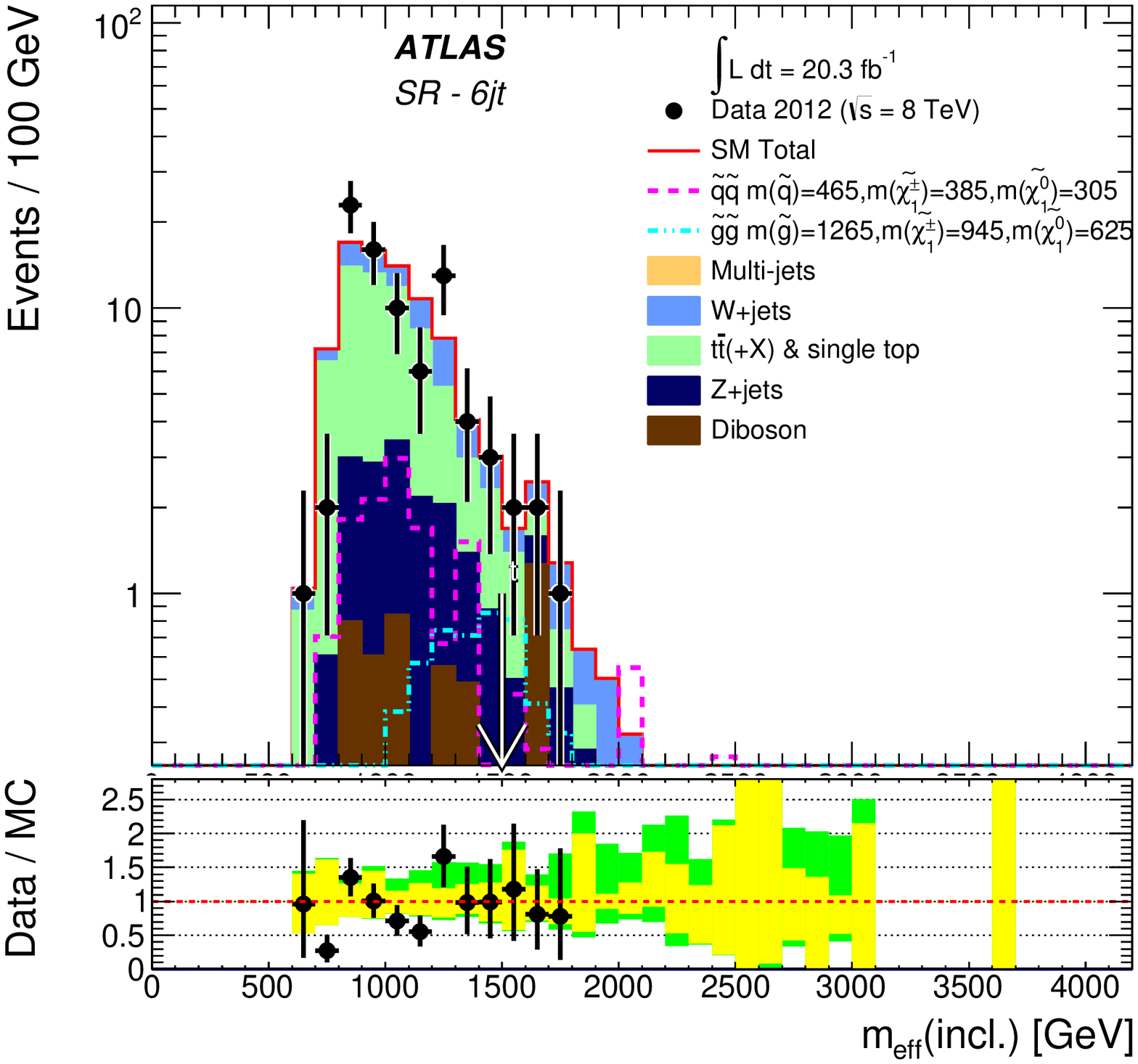}
\includegraphics[height=1.45in]{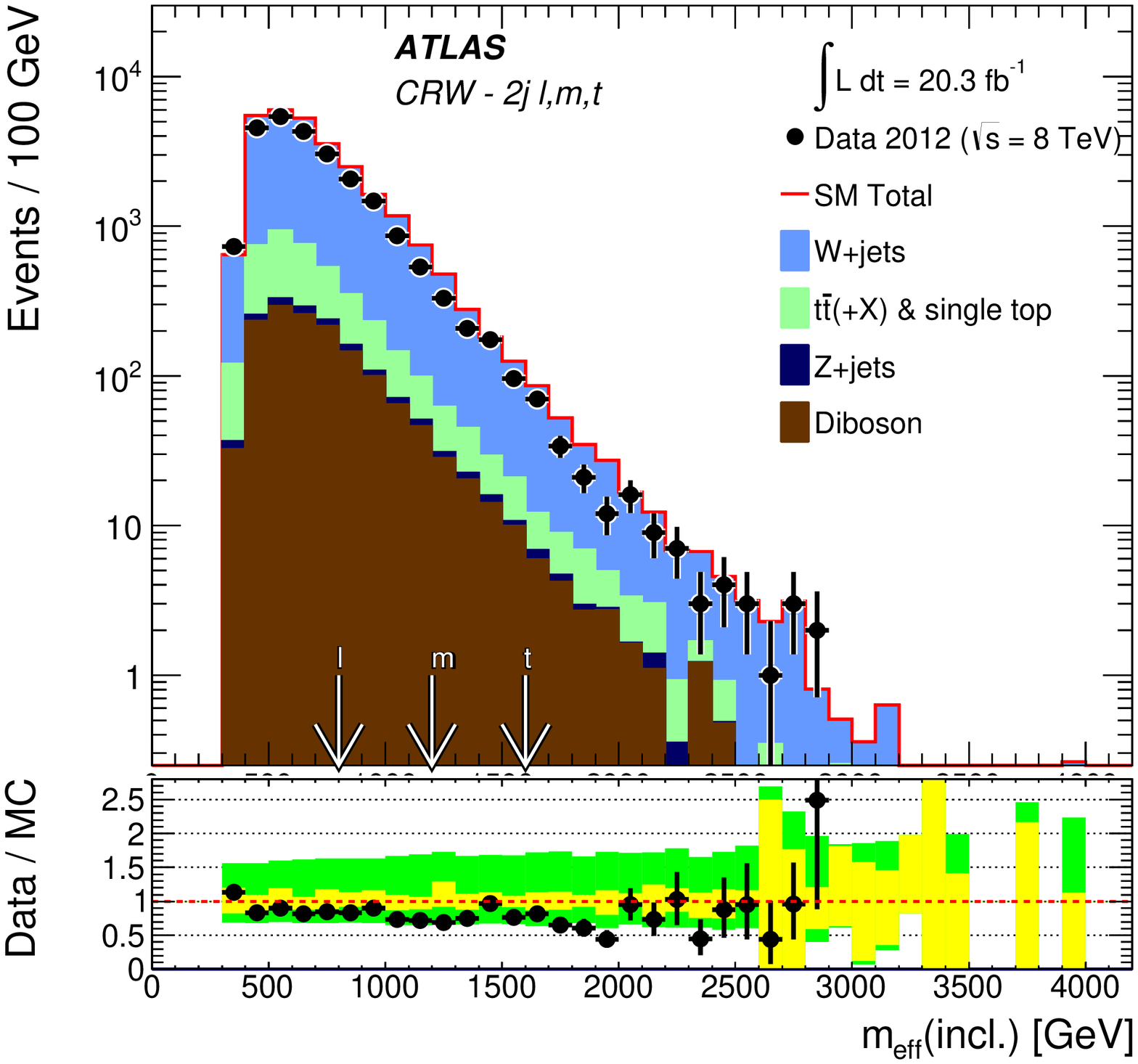}
\includegraphics[height=1.45in]{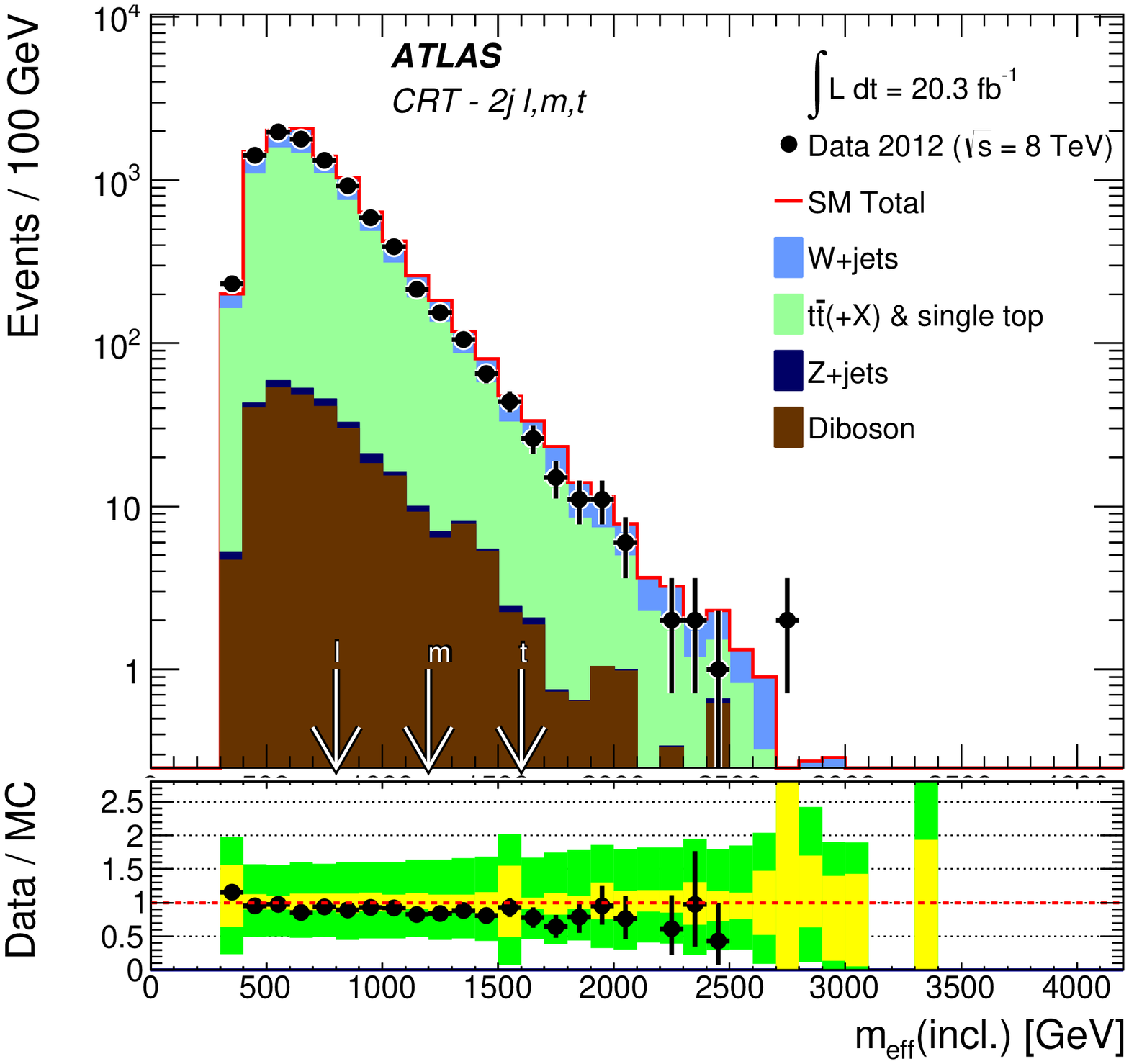}
\caption{Observed $m_{\mathrm{eff}}({\rm incl.})$ distributions for the 2-jet loose and 6-jet tight signal regions (two left plots) and in the 2j control regions CRW and CRT (two right plots)~\cite{0lep}.
}
\label{fig:SRs}
\end{figure}

This analysis uses 15 inclusive Signal Regions (SRs) depending on the minimal number of jets.
Several SRs may be defined for the same jet-multiplicity, distinguished by increasing background rejection.
The lower jet-multiplicity SRs focus on models characterised by squark pair production with short decay chains, while those requiring high jet-multiplicity are optimised for gluino pair production and/or long cascade decay chains.
SRs with at least 2 jets are dominated by W/Z+jets background, while SRs requiring at least 6 jets are dominated by $t\bar{t}$ (Figure~\ref{fig:SRs}).
Two dedicated SRs (2jW and 4jW) place additional requirements on the invariant masses of candidate W bosons (60 GeV $<m_W<$ 100 GeV) decaying to hadrons.

\section{Background estimation}

The dominant SM background processes that contribute to the event counts in the SRs are: $Z+$jets, $W+$jets, top quark pairs, single top quarks, and multiple jets.
To estimate them, four control regions (CRs) enriched in particular background sources are defined for each SR.
The CR definitions are given in Table \ref{tab:crdefs} and illustrated on Figure~\ref{fig:SRs}.
They are orthogonal to SRs and are optimised to have enough statistics and negligible expectation for the SUSY signal contamination, while minimising the systematic uncertainties arising from the extrapolation from the CR to the expectation in the SR.

The observed numbers of events in the CRs are used to obtain SM background estimates for the SR via a likelihood fit ~\cite{Cowan}.
The ratios of expected event counts from each background process between the SR and each CR enable observations in the CRs to be converted into background estimates in the SR using:
\begin{equation}
\label{eq:tf}
N\mathrm{(SR, scaled)} = N\mathrm{(CR, obs)} \times \left [\frac{N\mathrm{(SR, unscaled)}}{N\mathrm{(CR, unscaled)}}\right ],
\end{equation}
where $N$(SR, scaled) is the estimated background contribution to the SR by a given process, $N$(CR, obs) is the observed number of data events in the CR for the process, and $N$(SR, unscaled) and $N$(CR, unscaled) are a priori MC based estimates of the contributions from the process to the SR and CR, respectively.

\begin{table}
\footnotesize
\begin{center}\renewcommand\arraystretch{1.2}
\begin{tabular}{ l  c  c  c }
\hline
CR & SR background &  CR process & CR selection \\ \hline
CR$\gamma$ & $Z(\to\nu\nu)$+jets & $\gamma$+jets & Isolated photon \\
CRQ & Multi-jets & Multi-jets & SR with reversed requirements on (i) $\Delta\phi(jet, E_T^{\mathrm{miss}})_{min}$  \\
& & & and (ii) $E_T^{\mathrm{miss}}/m_{\mathrm{eff}}(N_{\rm j})$ or $E_T^{\mathrm{miss}}/\sqrt{H_{\rm T}}$\\
CRW & $W(\to\ell\nu)$+jets & $W(\to\ell\nu)$+jets & 30 GeV $<m_{\rm T}(\ell,E_T^{\mathrm{miss}}) < 100$ GeV, $b$-veto\\
CRT & $t\bar{t}$ and single-$t$ & $t\bar{t}\to b\bar{b}qq'\ell\nu$ & 30 GeV $<m_{\rm T}(\ell,E_T^{\mathrm{miss}}) < 100$ GeV, $b$-tag\phantom{o}\\ 
\hline
\end{tabular}
\end{center}
\caption{\label{tab:crdefs} Control regions used in the analysis.
}
\end{table}

\section{Results and interpretations}

\begin{figure}[htb]
\centering
\includegraphics[height=2in]{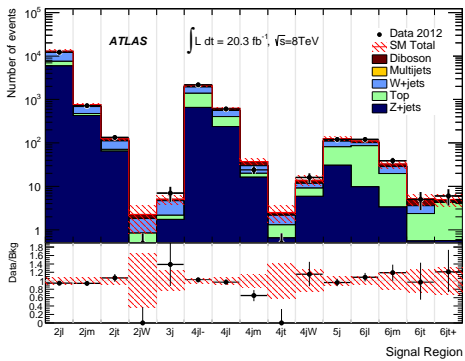}
\caption{\label{fig:PlotSR}
Comparison of the observed and expected event yields as a function of signal region~\cite{0lep}.}
\end{figure}

Cross-checks of the background estimates are performed using several Validation Regions (VRs) selected so to be distinct from the CRs, while maintaining a low probability of signal contamination.
Most VR observations lie within 1$\sigma$ of the background expectations.

\begin{figure}[htb]
\centering
\includegraphics[height=1.5in]{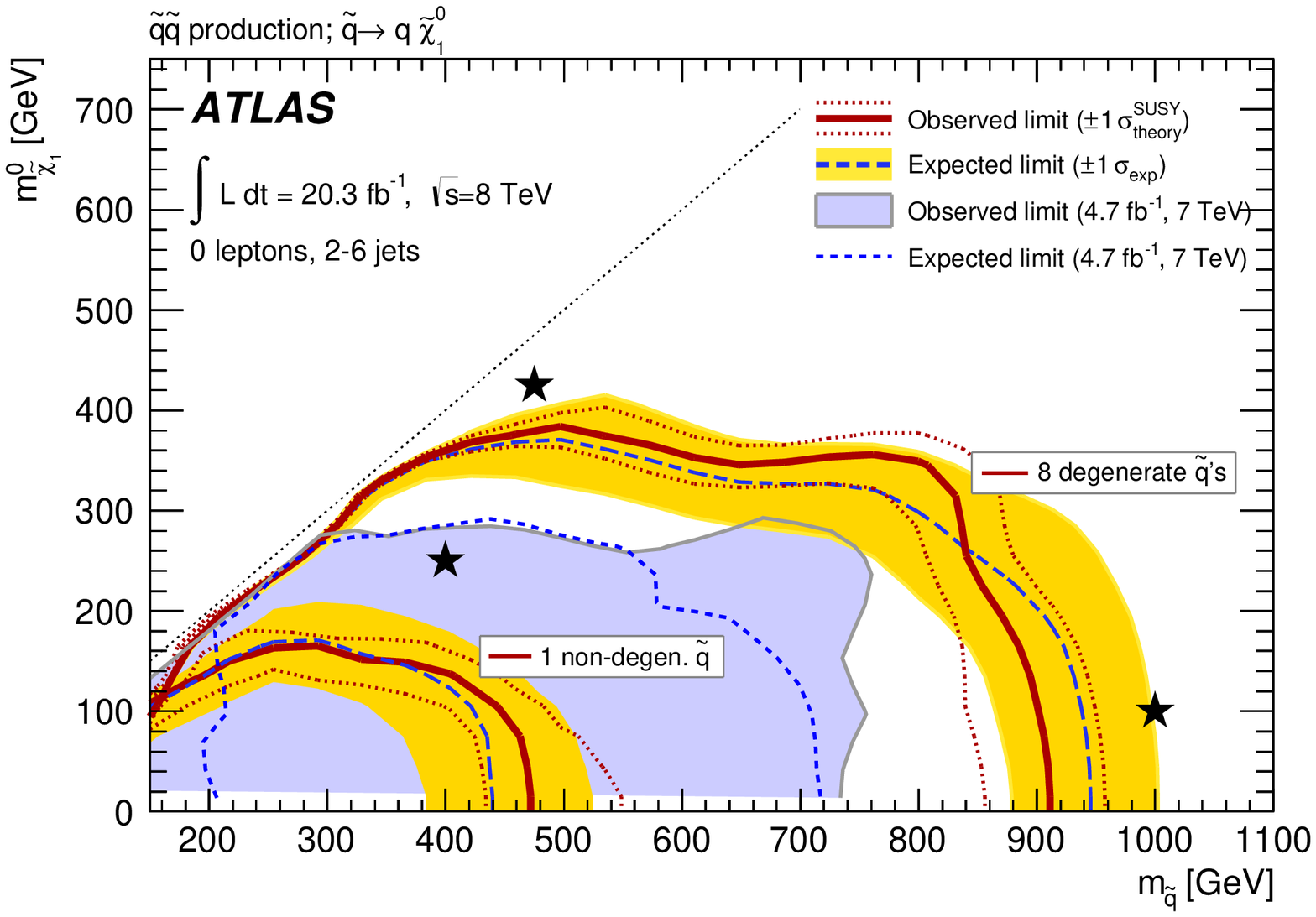}
\includegraphics[height=1.5in]{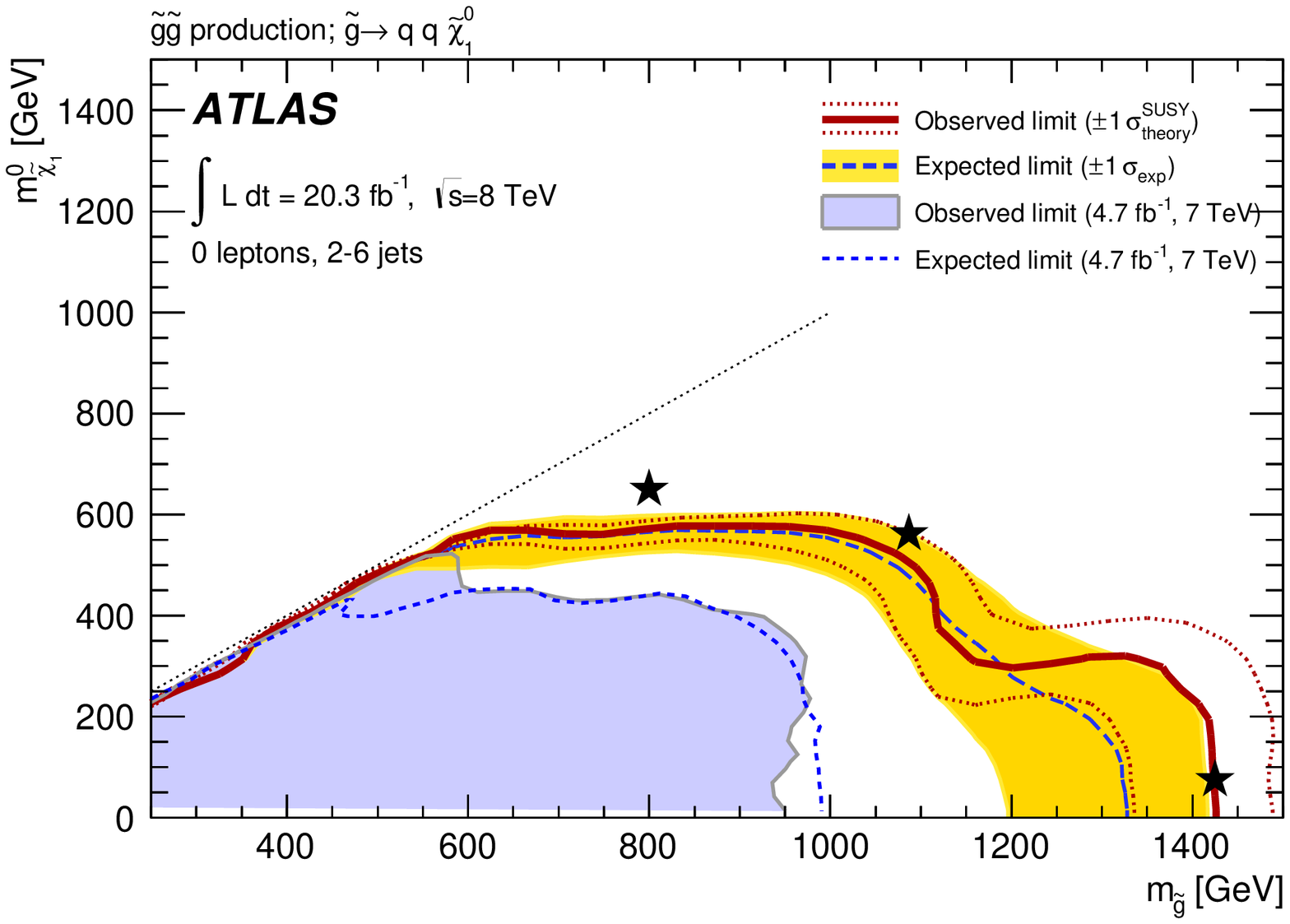}
\includegraphics[height=1.5in]{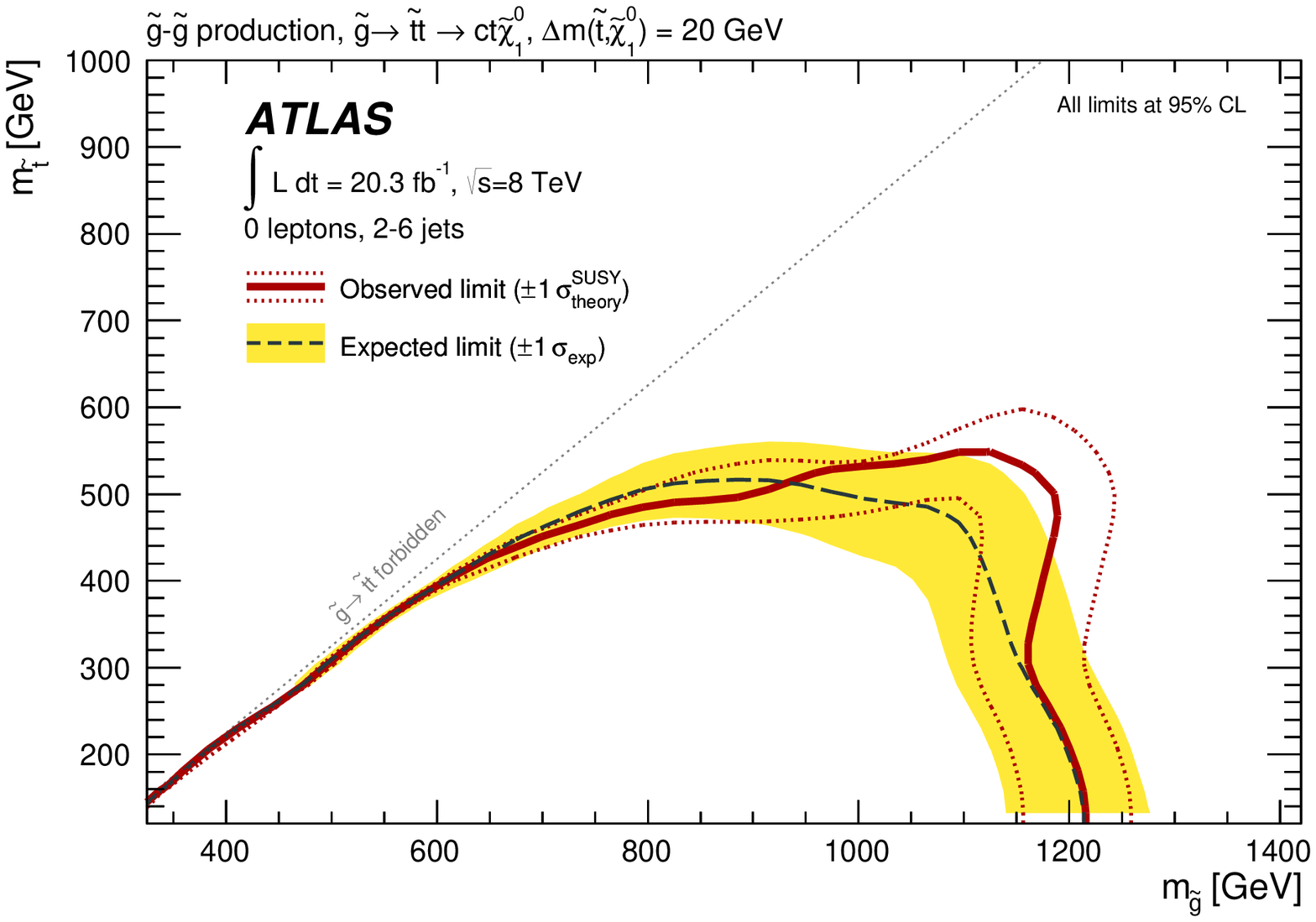}
\includegraphics[height=1.5in]{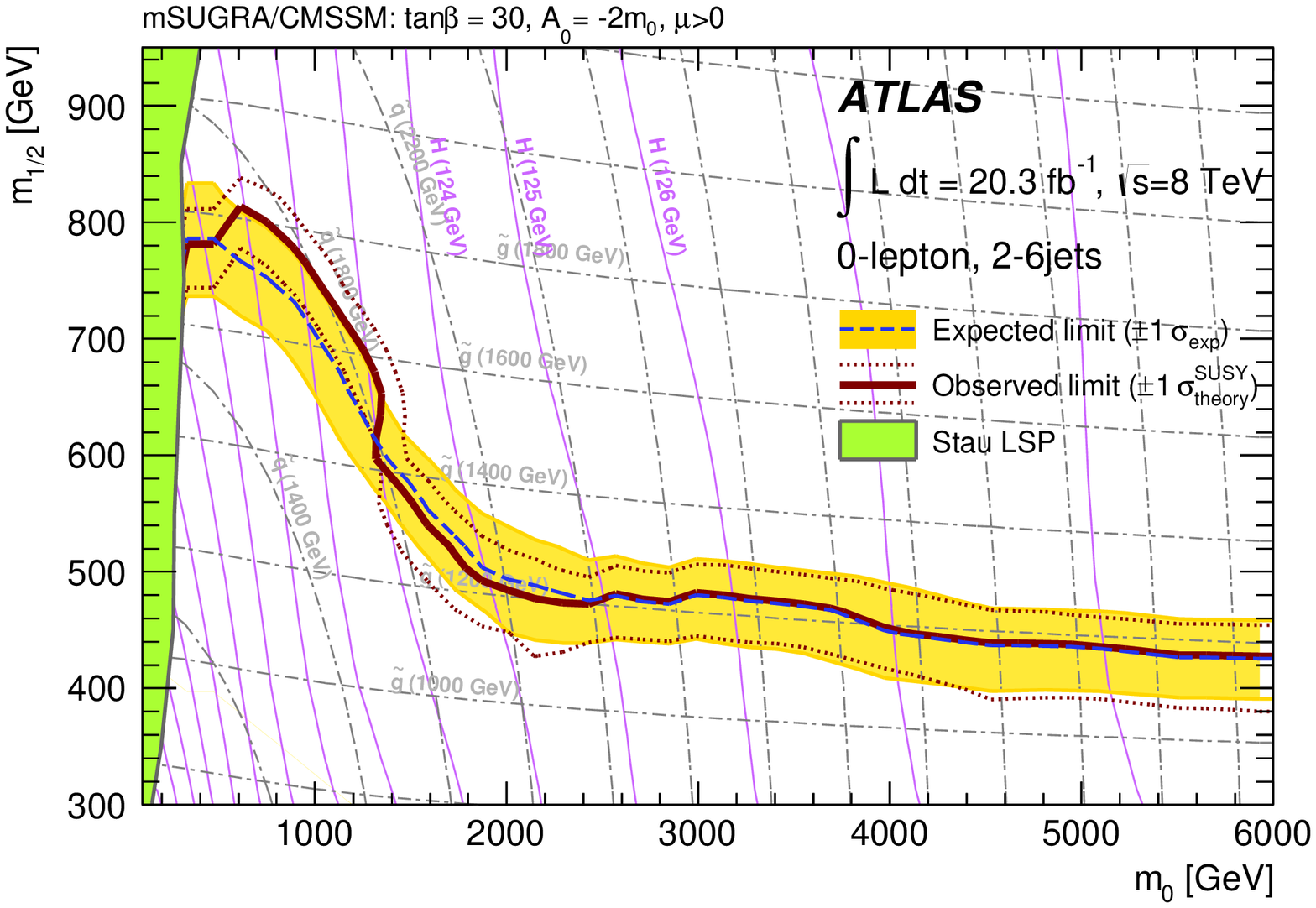}
\caption{ Exclusion limits for production of (top left) light-flavour squark pairs with decoupled gluinos, (top right) gluino pairs with decoupled squarks.
Exclusion limits for pair-produced gluinos each decaying into a $\tilde{t}$ and a $\tilde{\chi}_1^0$, with the subsequent decay $\tilde{t} \to c~ \tilde{\chi}_1^0$ and $\Delta m(\tilde{t},\tilde{\chi}_1^0)=20$ GeV (bottom left).
Exclusion limits for mSUGRA models presented in the $(m_0,m_{1/2})$-plane (bottom right)~\cite{0lep}.
}
\label{fig:Limit}
\end{figure}

There is no significant excess observed (Figure \ref{fig:PlotSR}), so the results are interpreted in terms of limits in several slices of the SUSY parameter space.
The combined limits were achieved choosing the best expected SR per model point.
Simplified models in which only production of gluino pairs or light-flavour squark pairs are considered and all other superpartners, except for the $\tilde{\chi}_1^0$ are decoupled thereby forcing each light-flavour squark or gluino to decay directly to one or more quarks and a $\tilde{\chi}_1^0$ are shown on Figure \ref{fig:Limit} (top).
Squark pair production limits are shown for 8 degenerate squarks, as well as for 1 non-degenerate squark (8 times lower cross-section).
When the $\tilde{\chi}_1^0$  is massless the limit on the gluino mass is 1330 GeV, and the limit on the squark mass is 850 GeV.
 For simplified models involving the pair production of gluinos, each decaying to a top squark and a top quark, with the top squark decaying to a charm quark and a $\tilde{\chi}_1^0$, the lower limit on the gluino mass extends to 1110~GeV for a top squark of mass 400~GeV (Figure \ref{fig:Limit} (bottom left)).
In mSUGRA models with $\tan\beta=30$, $A_0=-2m_0$ and $\mu>0$ (Figure \ref{fig:Limit} (bottom right)), squarks and gluinos of equal masses are excluded for masses below 1650 GeV.

\section{Conclusions}

The results of the search for squarks and gluinos in final states containing high$-p_T$ jets, $E_T^{\mathrm{miss}}$ and no electrons or muons, based on a 20.3 $fb^{-1}$ dataset of $\sqrt{s}=8$ TeV proton--proton collisions recorded by the ATLAS experiment at the LHC in 2012 are presented.
Good agreement is seen between the numbers of events observed in the data and the numbers of events expected from SM processes.
The results are interpreted in several SUSY models.
These results extend the region of SUSY parameter space excluded by previous searches with the ATLAS detector.



\begin{thebibliography}{99}


\bibitem{susy1} P. Fayet, Spontaneously Broken Supersymmetric Theories of Weak, Electromagnetic and Strong Interactions, Phys. Lett. B69 (1977) 489. 

\bibitem{susy2} G. R. Farrar and P. Fayet, Phenomenology of the Production, Decay, and Detection of New Hadronic States Associated with Supersymmetry, Phys. Lett. B76 (1978) 575Ð579. 

\bibitem{atlas} ATLAS Collaboration,
The ATLAS Experiment at the CERN Large Hadron Collider,
JINST 3 (2008) S08003.

\bibitem{0lep} ATLAS Collaboration,
Search for squarks and gluinos with the ATLAS detector in final states with jets and missing transverse momentum using $\sqrt{s}=8$ TeV proton-proton collision data,
[arXiv:1405.7875].

\bibitem{Cowan} G. Cowan, K. Cranmer, E. Gross, and O. Vitells,
Asymptotic formulae for likelihood-based tests of new physics,
Eur. Phys. J. C 71 (2011) 1554, [arXiv:1007.1727].

\end{thebibliography}
\end{document}